\documentclass{PoS}

\title{The QCD deconfinement critical point for $N_\tau=8$ with $N_f=2$ flavours of unimproved Wilson fermions}

\ShortTitle{The QCD deconfinement transition for $N_\tau=8$}

\author{\speaker{Christopher Czaban}, Owe Philipsen\\
        Institut für Theoretische Physik - Johann Wolfgang Goethe-Universität\\
        Max-von-Laue-Str. 1, 60438 Frankfurt am Main\\
		John von Neumann Institute for Computing (NIC)\\
		GSI, Planckstr. 1, 64291 Darmstadt, Germany \\
        E-mail: \email{czaban, philipsen@th.physik.uni-frankfurt.de} }

\abstract{
QCD at zero baryon density in the limit of infinite quark mass undergoes a first order deconfinement phase transition at a critical 
temperature $T_c$ corresponding to the breaking of the global centre symmetry.
In the presence of dynamical quarks this symmetry is explicitly broken. Lowering the quark mass the first order 
phase transition weakens and terminates in a second order Z(2) point. Beyond this line confined and deconfined regions are 
analytically connected by a crossover transition. 
As the continuum limit is approached (i.e. the lattice spacing is decreased) the region of first order transitions expands towards 
lower masses. We study the deconfinement critical point with standard Wilson fermions and $N_f=2$ flavours. 
To this end we simulate several kappa values on $N_\tau=8$ and various aspect ratios in order to extrapolate to the thermodynamic limit, 
applying finite size scaling. We estimate if and when a continuuum extrapolation is possible.
}

\FullConference{The 34nd International Symposium on Lattice Field Theory\\
		 24-30 July, 2016\\
		 University of Southampton, England}

\usepackage[utf8]{inputenc}
\usepackage[T1]{fontenc}
\usepackage{amsmath, amssymb}
\usepackage{caption}
\usepackage{dsfont}
\usepackage{booktabs}
\usepackage{cleveref}

\newcommand{\Tr}{\operatorname{Tr}}
\def\expect#1{\ensuremath{\left\langle{#1}\right\rangle}}
\newcommand{\abs}[1]{\lvert {#1}\,\rvert} 
\newcommand{\ee}{\ensuremath{\textrm{e}}}

\newcommand{\Loewe}{LOEWE-CSC}

\newcommand{\clqcd}{CL\kern-.25em\textsuperscript{2}QCD}

\newcommand{\kappacd}{\kappa_\text{crit}}

\begin{document}
\section{Introduction}
\begin{figure}[t]
    \centering
    \includegraphics[width=0.9\linewidth]{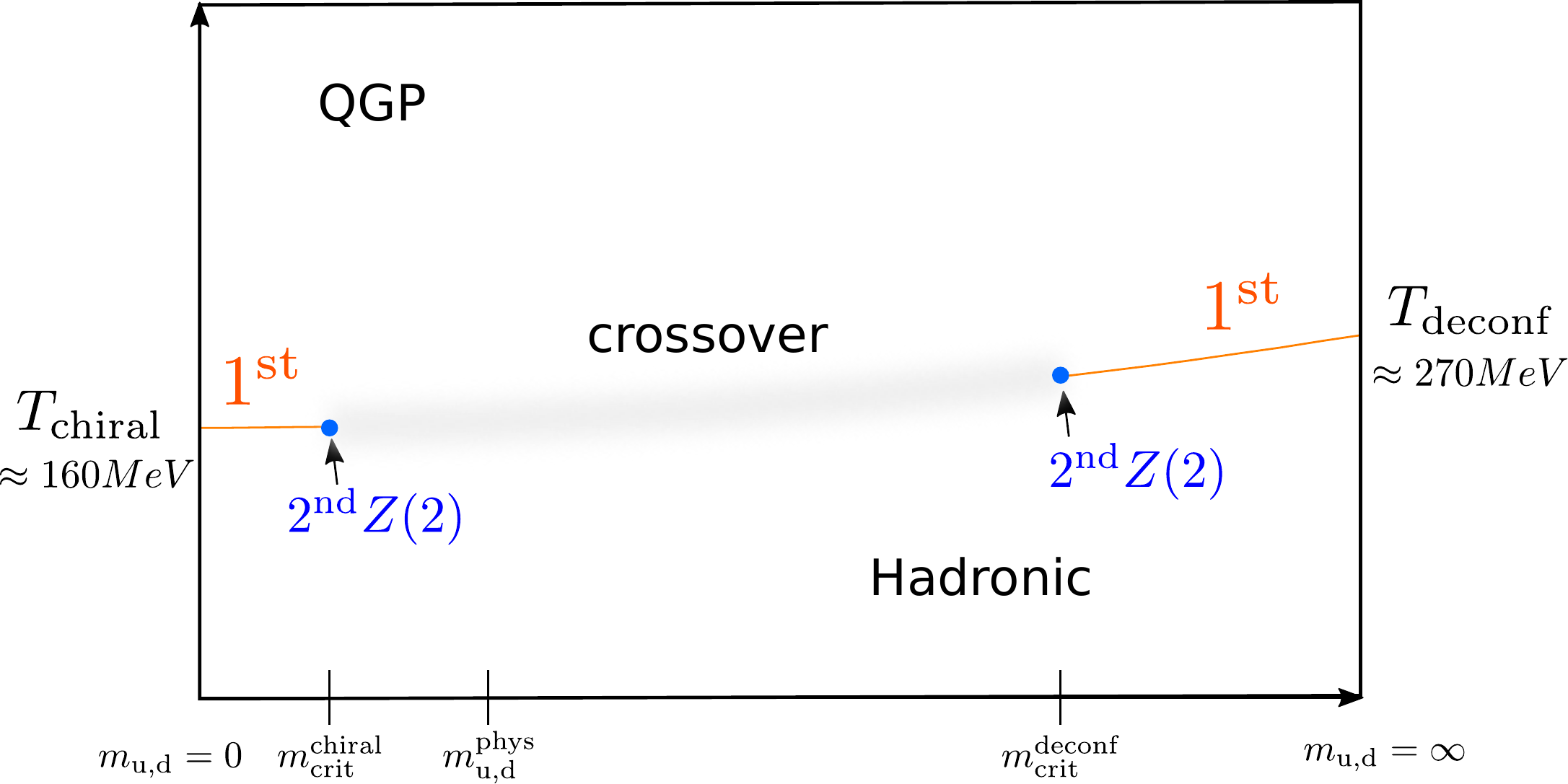}
    \caption{$N_f=2$ flavour temperature quark mass diagram: Schematic of a possible scenario of the nature of the phase transition as a function of the quark mass.
    There is a region of first order deconfinement phase transitions on the right side for heavy quark masses corresponding to the breaking of the centre symmmetry
    and another first order region on the left side for small quark masses associated to the breaking of the chiral symmetry. Inbetween there is an analytic crossover
    seperated from the first order regions by second order points belonging to the Z(2) universality class.}
    \label{fig:tmdiagram}
\end{figure}
\begin{figure}[t]
    \centering
    \includegraphics[width=0.90\linewidth]{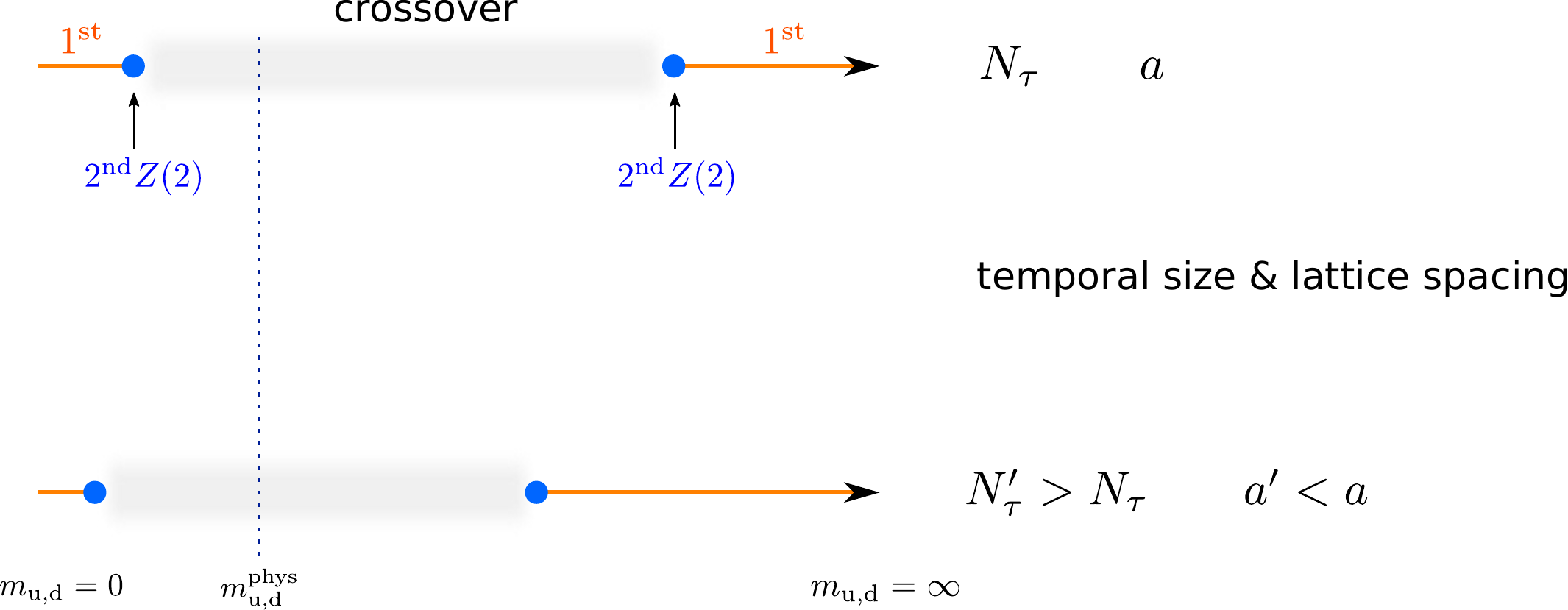}
    \caption{Schematic of the phase transitions on the $N_f=2$ flavour quark mass axis. As the lattice spacing $a$ is decreased, the
    region of first order phase transition expands in the heavy quark mass region and shrinks in the light quark mass region respectively.}
    \label{fig:phase_shift}
\end{figure}

During the last decades the phase structure of the QCD phase diagram has been extensively researched in many areas.
The region that can be investigated from first principles in lattice QCD is the phase structure at zero chemical 
potential which needs a solid 
understanding and consolidation to support research targeting finite $\mu$.
The nature of the phase transition at vanishing chemical potential depends on the quark mass $m_q$ and the number
of flavours $N_f$ under consideration. 
In the limit of infinite quark masses, i.e. excluding dynamical quarks, QCD has a first order deconfinement phase transition associated with
the spontaneous breaking of the centre symmetry at a critical temperature $T_c$. Going to finite quark masses, i.e. including dynamical quarks, breaks the centre symmetry
explicitly. For sufficiently large quark masses the phase transition remains first order but is weakend as the quark mass is decreased. 
At some critical value of the quark mass, here denoted as $m_\text{q,crit}^\text{deconf}$, it ends in a second order
point (the deconefinement critical point) belonging to the $Z(2)$ universality class. Beyond that point, i.e. for intermediate quark masses, the quark gluon plasma (QGP) and the hadronic phase
are analytically connected (c.f. fig.\ref{fig:tmdiagram}). At some small value of the quark mass $m_\text{q,crit}^\text{chiral}$ the 
phase transition becomes first order again. 
Additionally there is the dependence on the lattice spacing which affects the phase structure quantitatively.
These cut-off effects have been investigated in previous studies in which the $Z(2)$ 
transitions were observed to shift to smaller masses for $N_f=2, 2+1$ and $3$ flavours at zero and purely imaginary $\mu$
(c.f. \cite{deForcrand:2008zi,deForcrand:2007rq,Cuteri:2015qkq} and refs. therein).
The phase structure of the heavy quark mass region has been mapped out already (c.f. \cite{Saito:2011fs,Fromm:2011qi} and refs. therein)  but the studies took place on rather coarse lattices.  
Then of course the question arises which is the value of $m_\text{q,crit}^\text{deconf}$ in the continuum limit and if there is a consquence for the physical value of the quark mass. 
To answer the question an extrapolation to vanishing lattice spacing $a\rightarrow0$ is needed which due to the relation
$T_c=1/\left( a(\beta_c)N_\tau \right)$ requires to simulate at successively larger temporal lattice extents $N_\tau$ (c.f. fig.\ref{fig:phase_shift}).
By studying the QCD deconfinement phase transition on $N_\tau=8$ lattices this work presents a first step towards a continuum result of $m_\text{q,crit}^\text{deconf}$ for $N_f=2$ quark flavours.
Another important role in such a study of course plays the choice of the fermion discretization. 
In similar studies different fermion discretizations led to different results \cite{Iwasaki:1996ya,D'Elia:2005bv,Ejiri:2009ac}.
In order to understand how the deconfinement critical point explicitly moves as a function of $a$ we use the formulation of unimproved standard Wilson fermions.

\section{Simulation details} 
\label{sec:simulation_details}
In this work we use the standard Wilson gauge action
\begin{align}
    S_\text{gauge} = \beta\sum_{n}\sum_{\mu,\nu >\mu}\left( 1-\operatorname{Re}\Tr\left[ P_{\mu,\nu}(n) \right]  \right),
    \label{gauge_action}
\end{align}
with the lattice coupling $\beta=2N_c/g^2$ and Plaquette $P_{\mu\nu}(n)$ where $n$ denotes the lattice sites and $\mu,\nu$ are 
the Dirac indices. We simulate $N_f=2$ flavours of unimproved Wilson fermions with the standard Wilson action 
\begin{align}
    S_\text{fermion} = a^4 \sum_{N_f} \sum_{n,m} \bar\psi(n) D(n,m) \psi(m),
    \label{wilson_fermion_action}
\end{align}
where $D(n,m)$ denotes the Wilson fermion matrix
\begin{align}
    D(n,m) =& \;\delta_{nm} - \kappa\sum_{i=1}^{3}\left[ (1-\gamma_i)U_{\pm i}(n) \delta_{n+\hat i,m} \right] - \kappa\left[ (1-\gamma_0)\ee^{+a\mu}U_0(n)\delta_{m,n+\hat 0} \right. \\
    & \left. + (1+\gamma_0) \ee^{-a\mu}U_0^\dagger(m)\delta_{m,n-\bar 0}\right]. 
    \label{wilson_fermion_matrix}
\end{align}
The fermion mass is controlled via the hopping parameter $\kappa$ which is given by
\begin{align}
    \kappa = \frac{1}{2(am+4)}.
    \label{kappa}
\end{align}
The temperature can be tuned via the lattice coupling $\beta$ and is defined as
\begin{align}
    T=\frac{1}{a(\beta)N_\tau}.
    \label{eq:temperature}
\end{align}
In order to locate phase transitions we use the Polyakov loop,
\begin{align}
    L(n)=\frac{1}{3}\Tr_C\left[ \prod_{n_0=0}^{N_\tau-1}U_0(n_0,\bf{n}) \right],
    \label{polyakov_loop}
\end{align}
as order parameter and functions thereof. 
We fix the temporal lattice extent to $N_\tau=8$, the chemical potential to $\mu=0$ and scan for the critical $\kappacd$ 
at the second order $Z(2)$ point in the heavy quark mass region in $\kappa\in\left[ 0.1,\dots,0.13 \right]$. For each $\kappa$ value we use 3 
spatial lattice extents with a minimal aspect ratio of 4 in order to perform a finite size scaling study. The scans in temperature included 
3-4 $\beta$ values with $160k-800k$ HMC trajectories of unit length after $5k$ thermalization steps per $\beta$. For faster accumulation and better
control of statistics the runs were distributed on 4 Markov chains per $\beta$. The acceptance rate of the simulations was held at $\sim$ 75\%.
The simulations were performed with the OpenCL  
based code \clqcd\ \cite{Bach:2012iw} which is designed for running on graphic processing units (GPUs) on \Loewe\ \cite{LoeweRef} at Goethe university.
\section{Analysis}
The autocorrelation time on the Polyakov loop is computed with a python implementation of the Wolff method \cite{Wolff:2003sm}. Subsequently the data is binned appropriately to remove the
autocorrelation effects in functions of the observable. 
The main quantity for our analysis is 
\begin{align}
    B_n(x) = \frac{\expect{(\abs{L}-\expect{L})^n}}{\expect{(\abs{L}-\expect{L})^2}^\frac{n}{2}},
    \label{b_n_quantity}
\end{align}
with $x=\kappa,a\mu$ and $\mu=0$ in this work.  
In order to find $\kappacd$ we exploit properties of $B_3$, the skewness of the distribution and of $B_4$, 
the kurtosis of the distribution (trivially linked to the Binder cumulant \cite{Binder:1981sa} by substraction and multiplication of constants).
In a first step we locate the phase boundary $\beta=\beta_c(\kappa)$ for each $\kappa$ value by using $B_3(\beta_c)=0$.  
Ferrenberg-Swendsen reweighting \cite{FerrenbergSwendsenRef} is employed to
interpolate between the raw measurements of $B_3$.
Subsequently $B_4$ is evaluated at $\beta_c$ where it is a non-analytic step function and takes on particular values (see table \ref{tab:BinderValues})
according to the order of the phase transition.
Therefore $B_4$ is well suited to find the 
deconfinement critical point which marks the change from the first order to the crossover region. There $B_4$ takes on the value characteristic for the $Z(2)$ universality class.
On finite volumes $B_4$ is an analytic curve approaching the step function as the volume is increased. In the vicinity of the critical point 
$\kappacd$, $B_4$ is a function of $(\kappa-\kappa_\text{crit})N_\sigma^{1/\nu}$ only and can be expanded around $\kappa=\kappa_\text{crit}$ in a series to 
leading order (c.f. \cite{deForcrand:2010he})
\begin{align}
    B_4(\kappa,N_\sigma) = B_4(\kappacd,\infty) + b(\kappa-\kappacd)N_\sigma^{1/\nu}.
    \label{Eq:BinderCumulantExpansion}
\end{align}
Computing $B_4(\kappa,N_\sigma)$ for several $\kappa$ values on lattices with increasing spatial extents $N_\sigma$, curves with increasing slopes are obtained which close to the
thermodynamic limit altogether intersect at the universal value $B_4(\kappa_\text{crit},\infty)=1.604$ (see tab.\ref{tab:BinderValues}).
Then a finite size scaling study can be performed by fitting rel. (\ref{Eq:BinderCumulantExpansion}) to all $B_4$ data simultaneously which allows to extract $\kappacd$ 
at the intersection point.
To have a physical correspondence to $\kappa_\text{crit}$, the pion mass was computed using eight point sources per configuration. 
The lattice spacing was determined using a publicly available code described in ref. \cite{Borsanyi:2012zs} based on the Wilson flow method.
\begin{table}[t]
    \setlength{\tabcolsep}{7pt}
    \renewcommand{\arraystretch}{1.0}
    \centering
    \begin{tabular}{| c | c | c | c |}
        \hline
        & crossover & $1^\text{st}$ order & $2^\text{nd}$ order $Z(2)$  \\ 
        \hline
        $B_4$ & 3 & 1 & 1.604 \\
        $\nu$ & - & $1/3$ & 0.6301(4) \\
        $\gamma$ & - & 1 & 1.2372(5) \\
        \hline
    \end{tabular}
  \caption{Critical values of $\nu$, $\gamma$ and $B_4$
  		   for some universality classes}
  \label{tab:BinderValues}
\end{table}
\section{Numerical results}
\begin{figure}[t]
    \centering
    \includegraphics[width=1.0\linewidth]{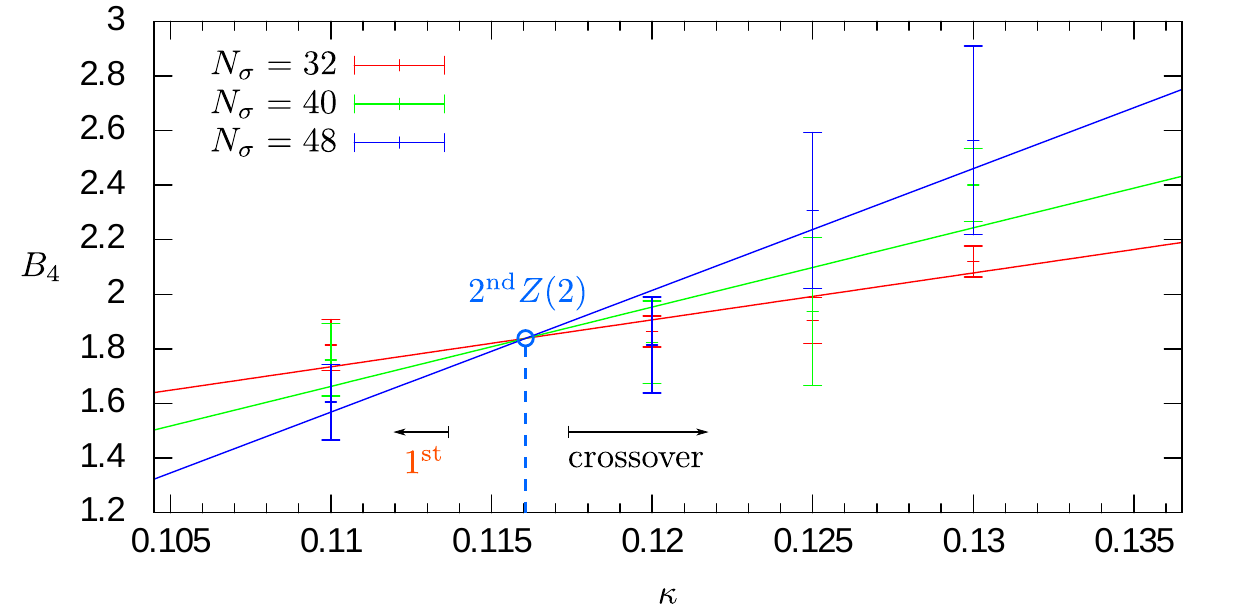}
    \caption[]{Fit of the expansion of $B_4$  given by rel.(\ref{Eq:BinderCumulantExpansion}) to the simulated 
    data according to section \ref{sec:simulation_details} 
    with a $\chi^2=0.93$ and $Q=48.58\%$. With a critical value of $\kappa_\text{crit}=0.1161(38)$ at the second order $Z(2)$ point, the universal value of $B_4=1.8387(984)$ and 
    the critical exponent $\nu=0.4251(1920)$.}
    \label{fig:results}
\end{figure}

The current status of the project for $N_\tau=8$ is depicted in fig.(\ref{fig:results}) which shows the fit of (\ref{Eq:BinderCumulantExpansion}) to the $B_4$ data generated
with the numerical setup described above. Due to a lack of statistics the errors on the data and the fitted parameters are still large and must be considered preliminary. 
The results for the critical deconfinement point, 
$B_4$ value at the intersection point and the critical exponent $\nu$ are 
$$\kappa_\text{crit}=0.1161\pm0.0038 \qquad B_4(\kappa_\text{crit})=1.8387\pm0.0984 \qquad \nu=0.4251\pm0.1920 .$$
The determined $B_4(\kappa_\text{crit})$ value is about $2.4\sigma$ away from the correct universal $B_4$ value, whereas the fitted critical exponent $\nu$ is only about 
$1.1\sigma$ away from its true value (c.f. table \ref{tab:BinderValues}). This confirms the observation in previous studies \cite{Cuteri:2015qkq} that the critical exponent is less prone
to finite volume effects compared to the $B_4$ quantity which has been observed to be larger in previous studies as well \cite{Cuteri:2015qkq,Philipsen:2014rpa}. 
The data points do not seem to describe curves with a constant slope but rather appear to flatten out towards smaller $\kappa$ values. This issue might be corrected by an increase of
the statistics. Another possible explanation which is very likely to apply are finite volume effects which is also reflected by the fact that the fitted value of $B_4$ is too large.
Another indication for this is that from the zero of $B_3$  slightly different values for $\beta_c$ were found for different aspect ratios for a given $\kappa$ value. 
This should not be the case in the thermodynamic limit.
In this case the situation worsens for smaller $\kappa$ values for the following reason:
As the $\kappa$ value is decreased the quark mass $m_q$ of the system is increased which implies a larger 
transition temperature $T_c$. Due to the relation $T_c=1/\left( a(\beta_c)N_\tau \right)$ this causes a smaller lattice spacing. Ultimately this results in a smaller physical volume
at smaller $\kappa$ values or larger quark masses $m_\text{u,d}$, respectively. The lattice spacing is computed for the largest
and smallest $\kappa$ value included in the fit and listed in table \ref{tab:BinderValues}. The difference between the physical volumes at the largest $\kappa=0.13$ and  
smallest $\kappa=0.11$ is about $19\%$. However, this is not a quantitative statement about how much more severe the finite volume effects are at $\kappa=0.11$ compared to $\kappa=0.13$.
Apparently the only way to reduce the finite volume effects is to increase the spatial extent which
increases the computational resources extensively in terms of simulation time. 
Advancing the study to lattices $N_\tau>8$ presents an even greater challenge in terms of computional effort.
Due to the reduced lattice spacing at larger $N_\tau$ the spatial lattice extent $N_\sigma$ should be increased to keep up the physical volume and avoid finite volume effects.
Another problematic point are the present cut off effects. The pion masses have been computed for the same $\kappa$ values listed in table \ref{tab:BinderValues}. 
At the largest $\kappa=0.13$ the bare pion mass does still not fulfill $1/m_\pi > a$ (see table \ref{tab:BinderValues}) which prevents a correct and 
meaningful measurement of the physical pion mass. 
Thus heavy quark effective theory (HQET) methods are needed to evaluate the pion masses (see \cite{Sommer:2015hea} for a recent review and references therein).
\begin{table}
    \setlength{\tabcolsep}{7pt}
    \renewcommand{\arraystretch}{1.5}
    \centering
    \begin{tabular}{| c | c | c | c | c | c |}
        \hline
        $\kappa$ & $\beta_\text{c}$ & $a$ [fm] & $am_\pi$ & $m_\pi$ [MeV] & $T_\text{c}$ [MeV] \\ 
        \hline
        0.1100 & 6.0303 & 0.0895(5) & 2.1310(6) & 4690(28) & 275(2)\\
    0.1300 & 5.9491 & 0.0947(6) & 1.3964(5) & 2904(17) & 260(2)\\
        \hline
    \end{tabular}
\end{table}

\section{Summary and Perspectives}

This study presents preliminary work done to determine the QCD deconfinement critical point at zero chemical potential in the heavy quark mass region for $N_f=2$ flavours.
Investigating the phase structue is a notoriously hard problem from a numerical viewpoint.
Among the difficulties to be faced the most severe ones are the successively smaller lattice spacing needed for 
the continuum extrapolation, the increasing lattice size necessary to suppress finite volume effects and the enormous amount of statistics needed to
properly sample the phases, especially in first order phase transition regions at larger volumes where tunneling gets suppressed.
From the trend of the data we believe to observe finite volume effects in the investigated parameter space, though larger statistics are needed in order to confirm this.
Apparently the lattice spacing, essentially governed by the temporal lattice extent and the transition temperature, is not yet sufficiently small in order 
to resolve the pion. Note that for $\kappa=0.13$ the
bare pion mass is $am_\pi\approx1.4$ which might indicate that on $N_\tau\geq10$ lattices the pion mass could already be sufficiently small to fullfil $1/m_\pi<a$. 
This will be clarified by follow-up simulations of this study.
However, HQET methods could be employed to determine the pion mass on lattices with smaller temporal lattice extents. 
In the near future advances in HPC technology, as well as improvements of the software \clqcd\ will render possible more efficient simulations at the relevant parameters.

\acknowledgments
This work is supported by the Helmholtz International Center for FAIR within the LOEWE
program of the State of Hesse. We thank the staff of LOEWE-CSC at GU-Frankfurt
for computer time and support as well as the NIC in Juelich for financial support.

\end{document}